\documentclass{PoS}

\title{Redshift measurement of the BL-Lac gamma-ray blazar PKS 1424+240}

\ShortTitle{Redshift measurement of PKS 1424+240}

\author{\speaker{A.C. Rovero}$^a$, C. Donzelli$^b$, A. Pichel$^a$ and H. Muriel$^b$\\
\llap{$^a$}Instituto de Astronom\'ia y F\'isica del Espacio (IAFE, CONICET-UBA), Buenos Aires, Argentina.\\
\llap{$^b$}Instituto de Investigaciones en Astronom\'ia Te\'orica y Experimental (CONICET) \& Observatorio Astron\'omico de C\'ordoba, C\'ordoba, Argentina.\\
E-mail: \email{rovero@iafe.uba.ar}} 

\abstract{
PKS 1424+240 is a BL-Lac blazar with unknown redshift detected at high-energy gamma rays by Fermi-LAT with a hard spectrum. It was first detected at very-high-energy by VERITAS and latter confirmed by MAGIC. Attempts to find limits on its redshift include three estimations by modeling gamma-ray observations, and one obtained by analyzing $Ly_{\beta}$ and $Ly_{\gamma}$ absorption lines observed in the far-UV spectra (from HST/COS) caused by absorbing gas along the line of sight. They allowed to constrain the redshift range to $0.6 < z < 1.19$, which places PKS 1424+240 in the very interesting condition to be one of the few candidates to be the most distant blazars detected at very-high-energy gamma rays.

Redshift determination of BL-Lac objects are difficult to achieve. We have found that redshift of blazars can be determined by its association to a galaxy group or cluster. To explore this possibility for PKS 1424+240, we have carried out spectroscopic measurements with the Gemini North telescope of galaxies in its field of view. In this work we present the optical spectrum of PKS 1424+240 and show preliminary results of the blazar environment characterization. Spectroscopic redshift using the optical spectrum of PKS 1424+240 could not be determined in this work.
}


\FullConference{The 34th International Cosmic Ray Conference,\\
		30 July- 6 August, 2015\\
		The Hague, The Netherlands}

\begin{document}

\section{Introduction}
The extragalactic background light (EBL) causes gamma-ray radiation in the range of very-high-energy (VHE; E>100 GeV) to be strongly attenuated by the photon-photon interaction. The EBL is a diffuse cosmological radiation field covering the UV to far-IR, it encompasses all radiative energy releases since recombination, and is dominated by formation of massive stars.
Gamma rays interact with low energy ambient EBL photons producing an electron-positron pair, which is the main attenuation effect for extragalactic VHE gamma-ray astronomy. As a consequence, all discovered extragalactic VHE sources are relatively close (typically $z <0.6$). At these redshifts, the attenuation effect for high-energy (HE; E$>$100 MeV) gamma rays is negligible. Thus, by modeling the drop in the spectral energy distribution (SED) from HE to VHE an estimation of the photon-photon interaction effects may be obtained, which in turn can be used to estimate either the spectral properties of the EBL (if the redshift is known) or the redshift of the source (e.g. \cite{aharonian06}, \cite{albert08}, \cite{finke09}, \cite{orr11}). For redshift estimations using this method a distribution of the EBL has to be assumed, and both the spectrum at HE and VHE have to be measured; the method is quite uncertain due to the indetermination of the EBL density and because to extended the SED from HE to VHE implies to make strong assumptions (e.g. \cite{dwek13}).

Extragalactic gamma-ray sources are mainly blazars, BL-Lacs and FSRQs. 
To measure the redshifts of BL-Lacs is a real challenge; the lack of emission and absorption lines in their spectra makes the determination of spectroscopic redshifts very difficult and often impossible (e.g. \cite{landt02}, \cite{sbarufatti05}) so, the method provided by gamma-ray astronomy mentioned above is used whenever possible.
In this sense, we have proposed an interesting alternative method to estimate the redshift of BL-Lac blazars in an indirect way. Given that BL-Lacs are typically hosted by elliptical galaxies, which in turn are associated to groups or clusters, we have proposed to realize spectroscopic observations to find the host group of galaxies associated with the blazar. The method was successfully proved with Gemini GMOS observations of the BL-Lac gamma-ray blazar PK 0447-439 \cite{muriel15}.\\

PKS 1424+240 is a BL-Lac blazar with unknown redshift detected at HE by Fermi-LAT with a hard spectrum \cite{abdo09}. It was first detected at VHE by VERITAS \cite{acciari10}, and latter confirmed by MAGIC \cite{aleksic14}. A $z \sim 0.24$ for the redshift of PKS 1424+240 was estimated using a statistical approach that correlates the drop in the gamma-ray spectra to real redshift measurements from known sources \cite{prandini11}, which strongly depends on the assumption that all blazars behave the same way and that there are not features in the gamma-ray spectra. Modeling the drop from HE to VHE of PKS 1424+240, an upper limit of $z < 0.66$ was found considering different EBL models \cite{acciari10}. Their result is in agreement with the upper limit of $z < 1.19$, for which only the lowest EBL estimation was used. A two-component synchrotron self-Compton model was found to describe the SED of the source well if it is located at $z \sim 0.6$ \cite{aleksic14}.
Leaving the gamma-ray domain, a photometric upper limit of $z < 1.11$ was reported \cite{rau12}. 
Also, a totally independent and more firm lower limit of $z > 0.6$ for the redshift of PKS 1424+240 was reported using recent UV observations taken with the HST/COS (which covers the range 1135-1795 \AA) \cite{furniss13}. They have analyzed the $Ly_{\beta}$ and $Ly_{\gamma}$ absorption lines observed in the far-UV spectra caused by absorbing gas along the line of sight. 
All these results place PKS 1424+240 in the very interesting condition to be one of the few most distant blazars detected at VHE, with redshift in a range never populated by other VHE blazar, i.e. $0.6 < z < 1.19$. This ambiguity in the redshift is still large enough to prevent precise studies of the EBL and the intrinsic blazar spectrum.\\

We submitted a proposal in September, 2014, to realize spectroscopic observation of PKS 1424+240 and galaxies in its environment with the Gemini North telescope in Hawaii. In this work we present the optical spectrum of PKS 1424+240 from data acquired using the GMOS-N spectrometer with high signal-to-noise ratio. We also present preliminary results of the blazar surrounding field characterization.

\section{Observations and data reduction}
Spectra of PKS 1424+240 and 30 other objects around it were obtained with the Gemini Multi-Object Spectrograph (GMOS), program GN-2015A-Q12 (PI: A.C. Rovero). A multislit mask was designed for this purpose using a 60 s exposure pre-image provided by Gemini, taken on March 15, 2015. The field covers $5\times5$ arcmin$^2$, with a pixel scale of 0.146 arcsec, which was centered on the position of PKS 1424+240. In order to characterize the blazar environment, extragalactic targets were selected in a way that maximizes the number of slits on the mask.

The grating in use was the B600$\pm$G5323 that has a ruling density of 600 lines/mm. Five exposures of 900 s each through a 1.0 arcsec slit were obtained with the central wavelengths of 540 nm, 550 nm, and 560 nm. Science targets have thus a total exposure time of 1.25 hours.

The spectroscopic data were acquired in queue mode on April 26, 2015. Observations were taken at airmass 1.25 with a seeing ranging from 0.64 to 0.80 arcsec. Flatfields, and the $CuAr$ lamp were also acquired to perform calibrations. A binning of $2\times2$ was used, yielding a scale of 0.146 arcsec per pixel and a theoretical dispersion of $\sim0.9$ \AA~ per pixel. Flux calibration was performed using the spectra of the standard star $Feige~ 66$.

All science and calibration files were retrieved from the Gemini Science Archive hosted by the Canadian Astronomy Data Center. The data reduction described below was carried out with the Gemini IRAF package. Flatfields were derived with the task {\sc gsflat} and the flatfield exposures. Spectra were reduced using {\sc gsreduce}, which does a standard data reduction, i.e. it performs bias, overscan, and cosmic rays removal, as well as the flatfielding derived with {\sc gsflat}. GMOS-North detectors are read with six amplifiers and generate files with three extensions. The task {\sc gmosaic} was used to generate data files with a single extension. The sky level was
removed interactively using the task {\sc gskysub} and the spectra were extracted using {\sc gsextract}. The sensitivity function of the instrument was derived using {\sc gsstandard} and the reference file for $Feige~ 66$ was
provided by the Gemini observatory. Science spectra were flux calibrated with {\sc gscalibrate} which uses the sensitivity function derived by {\sc gsstandard}.

\section{Results}
Figure 1 shows the new observed optical spectrum of PKS 1424+240 after data reduction and calibration, according to the procedure described in the previuos section. The spectrum covers the range 3940-6792 \AA. We have determined a signal to noise ratio for the continuum of $S/N \sim 300$ at 5500~\AA. 
There are not distinguishable features in the blazar spectrum other than local absorption lines. We clearly identify the Galactic Na~I absorption lines at 5891 \AA~ and 5898 \AA, and molecular Oxygen telluric absorption (see the insets of Figure 1).\\

\begin{figure*}[!ht]
  \centering
  \includegraphics[width=1.\textwidth]{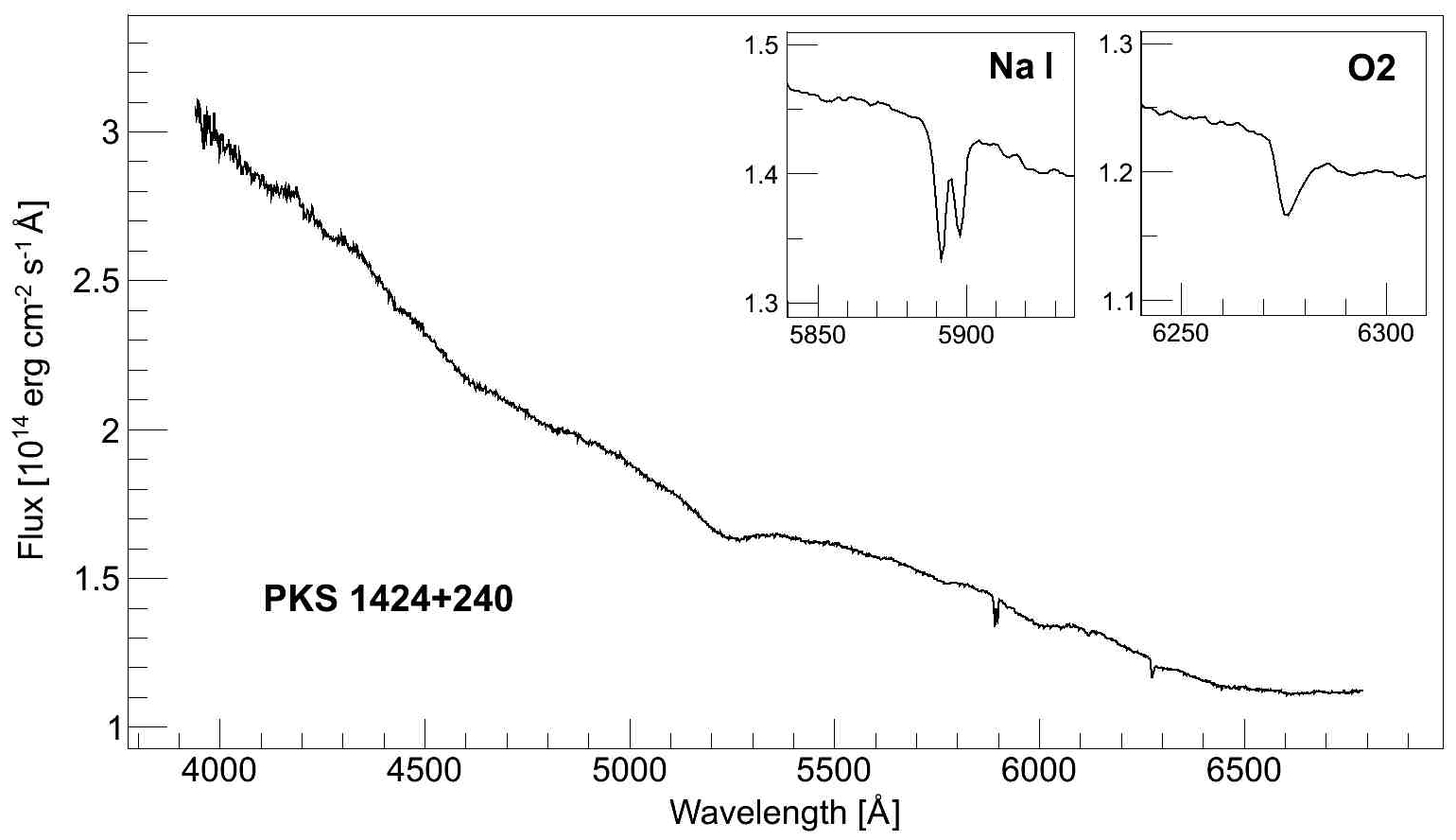}
  \caption{Observed optical spectrum of PKS 1424+240 with $S/N \sim 300$ at 5500 \AA. The insets show the Galactic Sodium lines and the molecular Oxygen telluric absorption.}
  \label{fig:espectro}
\end{figure*}

Table 1 shows the coordinates of the objects targeted for spectroscopy and their estimated magnitude from the $r'$ filter. The field we have observed is covered by the Sloan Digital Sky Survey (SDSS); all the objects in Table 1 are included in the SDSS photometric catalogue, and 3 of them are in the spectroscopic catalogue with a measured redshift: slit 8 (z$_{sdss} = 0.4688$), slit 13 (z$_{sdss} = 0.1212$), and slit 30 (z$_{sdss} = 0.1195$).
 
It has been suggested that BL-Lac spectra become featureless for jet-to-galaxy luminosity ratios $\gtrsim 10$ \cite{landt02}. They have simulated spectra from BL-Lacs for several luminosity ratios which suggest that for the case of PKS 1424+240 a jet-to-galaxy ratio $>40$ might be appropriate (comparing their spectra with that of Figure 1). Assuming that the host galaxy of our blazar is a bright elliptical, and taking the object luminosity as $r' = 14.56$ (from SDSS), we can estimate the minimum expected jet-to-galaxy luminosity ratio considering that PKS 1424+240 has a redshift $z > 0.6$. For an absolute magnitude $M_R \sim -23$ (a typical value for the host galaxy of BL-Lacs) we found a jet-to-galaxy ratio $> 100$, which is consistent with the featureless spectrum of PKS 1424+240.

\begin{table*}

\caption{Objects in the field of view with spectroscopic observations. Column 1: slit number; columns 2 and 3: RA and Dec (J2000.0); column 4: total $r'$ integrated magnitude. (\#) Has spectroscopic measurement at SDSS catalogue. (*) PKS 1424+240. (--) saturated.}
\center
\begin{tabular}{rccl|rccl}
\hline \hline
Slit  &RA (J2000.0) & DEC (J2000.0) &$m_{r'}$ & Slit  &RA (J2000.0) & DEC (J2000.0) &$m_{r'}$\\
\hline 
1  &  14:26:52.5 &  23:47:47.7  &  22.2 &  17 &  14:27:01.7 &  23:49:43.0  &  21.3 \\
2  &  14:26:53.8 &  23:47:18.5  &  22.6 &  18 &  14:27:03.4 &  23:49:10.0  &  21.6 \\
3  &  14:26:54.8 &  23:48:48.2  &  20.3 &  19 &  14:27:03.6 &  23:47:35.5  &  20.4 \\
4  &  14:26:56.4 &  23:50:29.5  &  21.5 &  20 &  14:27:05.2 &  23:49:17.9  &  21.6 \\
5  &  14:26:56.4 &  23:48:33.3  &  21.8 &  21 &  14:27:06.2 &  23:49:19.9  &  21.4 \\
6  &  14:26:57.6 &  23:48:09.7  &  19.1 &  22 &  14:27:06.5 &  23:47:14.2  &  21.0 \\
7  &  14:26:58.0 &  23:47:53.8  &  21.4 &  23 &  14:27:09.2 &  23:50:10.4  &  20.2 \\
(\#)~~8  &  14:26:59.0 &  23:47:42.0  &  20.5 &  24 &  14:27:10.4 &  23:49:40.1  &  19.8 \\
(*)~~9  &  14:27:00.4 &  23:48:00.4  &  ~--- &  25 &  14:27:10.9 &  23:49:37.5  &  19.8 \\
10 &  14:26:59.6 &  23:46:39.1  &  21.0 &  26 &  14:27:11.4 &  23:47:54.1  &  21.7 \\
11 &  14:27:01.7 &  23:48:33.1  &  18.8 &  27 &  14:27:11.6 &  23:47:34.5  &  21.9 \\
12 &  14:26:59.4 &  23:50:20.0  &  20.6 &  28 &  14:27:12.6 &  23:46:16.1  &  22.0 \\
(\#) 13 &  14:27:01.8 &  23:46:31.1  &  17.5 &  29 &  14:27:14.3 &  23:48:13.1  &  18.4 \\
14 &  14:27:04.1 &  23:47:49.8  &  20.1 &  (\#) 30 &  14:27:14.5 &  23:50:07.6  &  17.2 \\
15 &  14:26:54.6 &  23:50:32.6  &  21.5 &  31 &  14:27:12.8 &  23:48:50.9  &  22.3 \\
16 &  14:27:00.4 &  23:49:38.4  &  21.6 & & & & \\
\hline
\end{tabular}

\label{tab:data}
\end{table*}


\section{Searching for redshift determination}
Within the range of wavelength covered by the new optical spectrum of PKS 1424+240 we have not identified any
spectral line from any chemical element other that those interpreted as either Galactic or terrestrial. Alternatively, we checked visually possible evidences of the presence of ten absorption and six known emission lines with rest frame wavelengths in the range 2800 \AA~to 6800 \AA. This was done in the redshift range $z=0.0-1.5$. No significant or marginal evidence of coincidences between these commonly observed lines and the spectrum of PKS 1424+240 was found. Consequently, no spectroscopic estimation for the redshift of this blazar could be determined in this work.

\section{Summary}
We observed the BL-Lac blazar PKS 1424+240 with the Gemini North telescope on April 26, 2015, to perform spectroscopic measurements with a signal to noise ratio of $S/N = 300$ at 5500 \AA. We were not able to identify any spectral feature to obtain a spectroscopic redshift of this blazar; only Galactic and telluric spectral lines were identified (Na~I and O2). A jet-to-galaxy luminosity ratio was estimated to be $> 100$, which is consistent with the featureless kind of spectrum of PKS 1424+240. We have also obtained optical spectroscopic data on 30 other objects to characterize the blazar environment. We are currently analyzing those spectra searching for the presence of a group or cluster of galaxies to which associate PKS 1424+240. Three out of the 30 objects have redshift measurements at the SDSS spectroscopic catalogue to which we can compare our estimations.

\acknowledgments {This work is based on observations obtained at the Gemini Observatory, which is operated by the Association of Universities for Research in Astronomy, Inc., under a cooperative agreement with the NSF on behalf of the Gemini partnership: the National Science Foundation (United States), the National Research Council (Canada), CONICYT (Chile), the Australian Research Council (Australia), Minist\'{e}rio da Ci\^{e}ncia, Tecnologia e Inova\c{c}\~{a}o (Brazil) and Ministerio de Ciencia, Tecnolog\'{i}a e Innovaci\'{o}n Productiva (Argentina). The authors are all members of ``Carrera del Investigador Cient\'ifico'' of CONICET.}

\end{document}